\documentstyle[multicol,aps,epsfig]{revtex}

\begin{document}
\draft
\preprint{Submitted to Phys. Rev. B R.C.}

\title{Dynamical Phases  of Driven Vortices Interacting with
 Periodic Pinning}  

\author{ Gilson Carneiro$^{*}$}
\address{Instituto de F\'{\i}sica\\Universidade Federal do Rio de Janeiro\\  
C.P. 68528\\ 21945-970, Rio de Janeiro-RJ \\ Brasil }
\date{\today}
\maketitle
\begin{abstract}
The finite temperature dynamical phases of vortices  in films driven by a
uniform force and interacting with  the periodic pinning potential
of a square lattice of columnar defects are investigated
by Langevin dynamics simulations of a London model. 
Vortices driven along the [0,1] direction and at densities for which 
there are more vortices than columnar defects ($B>B_{\phi}$)
are considered. At low temperatures, two new dynamical phases, elastic
flow and plastic flow, and a sharp transition between them are identified  
and characterized according to the behavior of the vortex spatial
order, velocity distribution and frequency-dependent velocity correlations.

\end{abstract}

\begin{multicols}{2}

There is a great deal of current interest in the study of dynamical
phases of vortices driven by an external
force and interacting with various arrays of pinning centers.
One possibility is periodic pinning resulting from a lattice of
artificial defects. Understanding the dynamical phases 
in this case is of  interest because 
it may be possible to  observe them in
superconducting  films with periodic arrays of  holes, magnetic  dots
and columnar defects (CD). Several techniques have been developed  to
fabricated these films and studies of vortex dynamics in them have been
reported \cite{lyk,mat}. Theoretical workers have 
carried out investigations, mostly numerical, of driven vortices under
periodic pinning \cite{marti,norea,norb,dom}. However, as discussed here,
several questions remain open. 

To be specific, this paper considers two-dimensional 
vortices interacting  with a columnar defect  lattice (CDL).
The motion of $N_v$ such vortices is assumed to be 
governed by Langevin equations for massless particles, which for the
$l$-th vortex reads\cite{ehb}, 
\begin{equation}
\eta \frac{d{\bf r}_l}{dt}= {\bf F}^{vv}_l + {\bf F}^{v-cd}_l
+  {\bf F}_{d} + {\bf \Gamma}_l\;, 
\label{eq.lan}  
\end{equation}
where $\eta$ is the friction coefficient,
 ${\bf F}^{vv}_l=-\sum^{N_v}_{j\neq l=1}{\bf \nabla}_lU^{vv}
({\bf r}_l-{\bf r}_j)$ is the force of interaction with other vortices,  
${\bf F}^{v-cd}_l=- \sum_{\bf R}{\bf \nabla}_l U^{v-cd} ({\bf
r}_l-{\bf R})$ is the force of interaction with the CDL, 
${\bf R}$ denotes the CDL positions,  
${\bf F}_{d}$ is the driving force,  and
${\bf \Gamma}_l$ is the random force appropriate for  temperature $T$.

In the absence of  driving (${\bf F}_{d}=0$), the equilibrium phases of
this system ( hereafter called
zero-drive state) are studied (at low $T$) in Refs.\cite{blam,nor,wmgc}.
 
For ${\bf F}_{d}\neq 0$ the simplest situation occurs at very high
drives \cite{hdp}.  In this case the vortices center of mass (CM) move with
velocity ${\bf V}_{d}={\bf F}_{d}/\eta$ and in the CM frame of
reference, defined by ${\bf r}^{\prime}_l={\bf r}_l-{\bf V}_dt$,
 the equations of motion are as in Eqs.\ (\ref{eq.lan}), but without
the driving term ${\bf F}_{d}$ and with the vortex-CDL force replaced by the
time-dependent interaction ${\bf F}^{\prime(v-cd)}_l(t)=\sum_{\bf
Q}(-i{\bf Q}) U^{v-cd}({\bf Q}) e^{i{\bf Q}\cdot {\bf r}^{\prime}_l}
e^{i{\bf Q}\cdot {\bf V}_dt}$, where ${\bf Q}$ denotes the
CDL reciprocal lattice vectors and $U^{v-cd}({\bf Q})$ is the Fourier
transform  of $U^{v-cd}({\bf r})$. For high enough drives the
Fourier components in ${\bf F}^{\prime(v-cd)}_l(t)$ for which ${\bf
Q}\cdot {\bf V}_d\neq 0$ oscillate fast, thus having a negligible effect on
the vortex trajectory \cite{hdp}. The vortex-CDL force in the CM frame reduces
then  to the static one  obtained by summing the Fourier
components in ${\bf F}^{\prime(v-cd)}_l(t)$ with ${\bf Q}\cdot {\bf V}_d =
0$ (or ${\bf Q}\perp {\bf V}_d$). This force is just that resulting
from the  potential obtained by  averaging $U^{v-cd} ({\bf r})$ in the
direction of drive. Thermal fluctuations lead, after sufficient time,
to relaxation into the thermodynamic equilibrium state of the $N_v$
vortices interacting between themselves and with this average potential.
This state is referred to here as the infinite-drive state.

Since at low $T$ the infinite-drive  and the zero-drive states are generally
different, reordering of the vortices 
must take place  as the driving force is varied. This predicts the
existence of at least two dynamical phases and,  possibly, of  
one  dynamical phase transition. One novel result obtained in this
paper is to show that this prediction is correct. Studies of vortices
interacting with a CDL carried out at $T=0$  find a rich variety of
phases, but no reordering to the infinite drive-state \cite{norea,norb}.    

For driven vortices interacting with a
random pinning potential, the existence of reordering and of a dynamical
 transition  as the driving force is varied are well established
\cite{rpin}. In this case the pinning potential averaged in the
direction of drive remains random in the direction perpendicular to it,
and  has non-trivial effects on the dynamical phase diagram, as extensively
discussed in Refs.\cite{giam}.  
For motion on a periodic pinning potential, numerical simulations
reported in Refs.\cite{gmc} show that vortex reordering to the
 infinite-drive state only occurs if sufficient
thermal fluctuations are present. Otherwise the vortices are trapped
in a metastable state.

The numerical simulations reported here investigate the properties of
the moving vortices steady-states as a function of 
the driving force, for a range of $T$  beginning at a value 
well below the  infinite-drive state melting temperature,
$T_m$, and extending up to $T_m$. The simulations are initialized
with the vortices in the infinite-drive state and  large enough $F_d$.
In subsequent runs, at the same $T$, $F_d$ is  progressively decreased.
The main conclusions  reached by this approach  are  that for
$T<T_m$ a dynamical phase    
with the spatial symmetry of the of the infinite-drive state, elastic flow
and long-range time order exists  for
$F_d>F_c(T)$. At $F_d=F_c(T)$  a  transition separates it from another
dynamical phase with distinct spatial order and plastic flow. It is
found that the transition is  sharp at low $T$ and that $F_c(T)$ 
increases as  $T\rightarrow T_m$.

The details of the model are as follows \cite{gmc}. 
Vortices and CD are   placed on a square lattice subjected to periodic 
boundary conditions (hereafter called the space lattice),
with $N$ square primitive cell of dimensions $d\times d$, oriented with the
sides  in the $x$ and $y$-directions. The vortex has a core of
linear dimension $d_v>d$ ($d_v\sim 2\xi(0)$). The interactions between
vortices and between vortices and CD are chosen to model films in the
London limit.  The vortex-vortex interaction potential is  a screened
Coulomb one \cite{min}, appropriate for the space lattice, with a short 
distance cutoff to account for the vortex core \cite{ehb}. 
That is, $U^{vv}({\bf r})$ is the lattice Fourier transform of
$U^{vv}({\bf k})= 4\pi^2 J\exp{(-\kappa^2/\kappa^2_c)}/ 
(\kappa^2+\Lambda^{-2})$, where $J=(\phi^2_0d_v/32\pi^2\lambda^2)$ is
the energy scale for vortex-vortex interactions \cite{min}, $\lambda$ is the
penetration depth, $\kappa^2= 4\sin^2{(k_xd/2)}+4\sin^2{(k_yd/2)}$,
$\kappa_c=2\sin{(\pi d/2d_v)}$ is the cutoff in k-space, and $\Lambda$ is
the screening length ($\Lambda>\lambda$).     
The CD-lattice has $N_{cd}$ sites arranged on a square
lattice, with lattice constant $a_{cd}$, and is 
commensurate with the space lattice. The  vortex-CD potential  is
chosen with depth $U^{v-cd}(0)$, range
$R_{cd}>d_v$ and a spatial dependence that gives square
equipotentials and an attractive  pinning force of constant modulus $F_p=\mid
U^{v-cd}(0)\mid /R_{cd}$. That is, $U^{v-cd}({\bf
r})=U^{v-cd}(0)(1-\mid x\mid/R_{cd})$ for $\mid x \mid\leq R_{cd}$,
$-\mid x\mid \leq y \leq \mid x\mid$, and $U^{v-cd}({\bf
r})=U^{v-cd}(0)(1-\mid y\mid/R_{cd})$ for $\mid y \mid\leq R_{cd}$,
$-\mid y \mid \leq x \leq \mid y \mid$.

The algorithm for the simulation of this model follows the usual
procedure\cite{lsi}. In the results reported next space lattices with
$N=256\times 256$ and $512\times 512$ sites are used to accommodate a CDL
with $N_{cd}=64$ and $256$, respectively, both with $a_{cd}=32d$. 
Other parameters are fixed at $d_v=4d$, $\Lambda=160d$,
$U^{v-cd}(0)=-J$,  $R_{cd}=3d_v=12d$, from which it follows that $F_p=J/12d$.
The vortex systems studied have $N_v=1.25N_{cd}$ and $2N_{cd}$ or, in
terms of the magnetic induction, $B=1.25B_{\phi}$ and $B=2B_{\phi}$,
where $B_{\phi}=\phi_0/a^2_{cd}$ is the matching field. Typical run
times are  $2-5\times 10^6 $ steps  of $10^{-2}\tau$ ($\tau=\eta d^2/J$
is the unit of time), after the steady state is reached. The reported results represent the average over several  different realizations of the random force($\sim 10$).
The driving force is assumed to be  along $y$ (CDL (0,1)-direction).

\begin{figure}
\epsfig{file=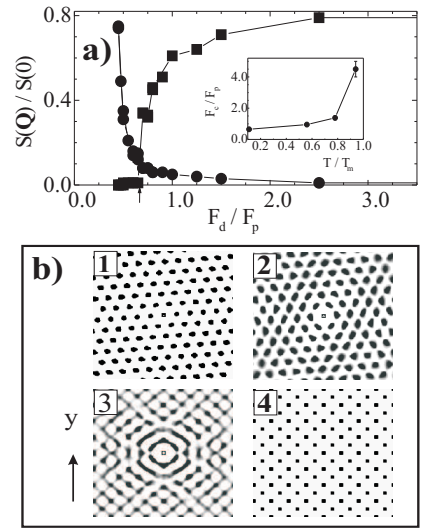,height=7cm,width=5.44cm,clip=}
\vspace{0.2cm}
\caption{a) Structure function for $B=2B_{\phi}$ and $T/T_m=0.1$. Squares
(circles): ${\bf k}$= typical infinite (zero)- drive state reciprocal-lattice vector of smallest modulus. Inset $F_c$ vs. $T$. b) Corresponding $P({\bf r})$ for: 1) infinite-drive state,
2) elastic flow phase at $F_d=0.7F_p$, 3) plastic flow phase at $F_d=0.6F_p$, 4) zero-drive  state for $F_d<0.45F_p$. The origin ${\bf r}=0$ is at each picture center.  Gray scale in each picture is proportional to $P({\bf r})$, but scales in different pictures are unrelated.}    
\label{fig:f1nn}
\end{figure}

To characterize the  dynamical phases  the following quantities 
are calculated in the steady-state regime: i) Time
average of the individual vortex velocities, ${\bf v}_j$ 
($j=1,...,N_v$),  their mean or the center of mass (CM) velocity, 
${\bf V}^{cm}=\sum_j{\bf v}_j/N_v$, and root-mean square deviation 
$\Delta V^{cm}_{\alpha} = 
\sqrt{\sum_j(v_{j\alpha}-V^{cm}_{\alpha})^2/N_v}$, 
where $\alpha = x,y$. ii)
Time-averaged density-density correlation function, $P({\bf r})$, and
its Fourier transform, the structure function, $S({\bf k})$. 
iii) Time-dependent correlation functions for
the CM velocity, 
$C^{cm}_{\alpha}(t)=\langle V^{cm}_{\alpha}(t+s)V^{cm}_{\alpha}(s)\rangle_s$, 
and for the vortex
velocity autocorrelation function averaged over all vortices, 
$C^{sf}_{\alpha}(t)=
\frac{1}{N_v}\sum_i \langle v_{i\alpha}(t+s)v_{i\alpha}(s)\rangle_s $,
where $\langle \rangle_s$ denotes average with respect to time $s$,
and the respective Fourier transforms $C^{cm}_{\alpha}(\omega)$ and 
$C^{sf}_{\alpha}(\omega)$. 
Physically,  $P({\bf r})$ is proportional to the 
probability of finding a pair of vortices separated by ${\bf r}$,
whereas  $C^{cm}_{\alpha}(\omega)$  is proportional to the noise power
spectrum.

\begin{figure}
\epsfig{file=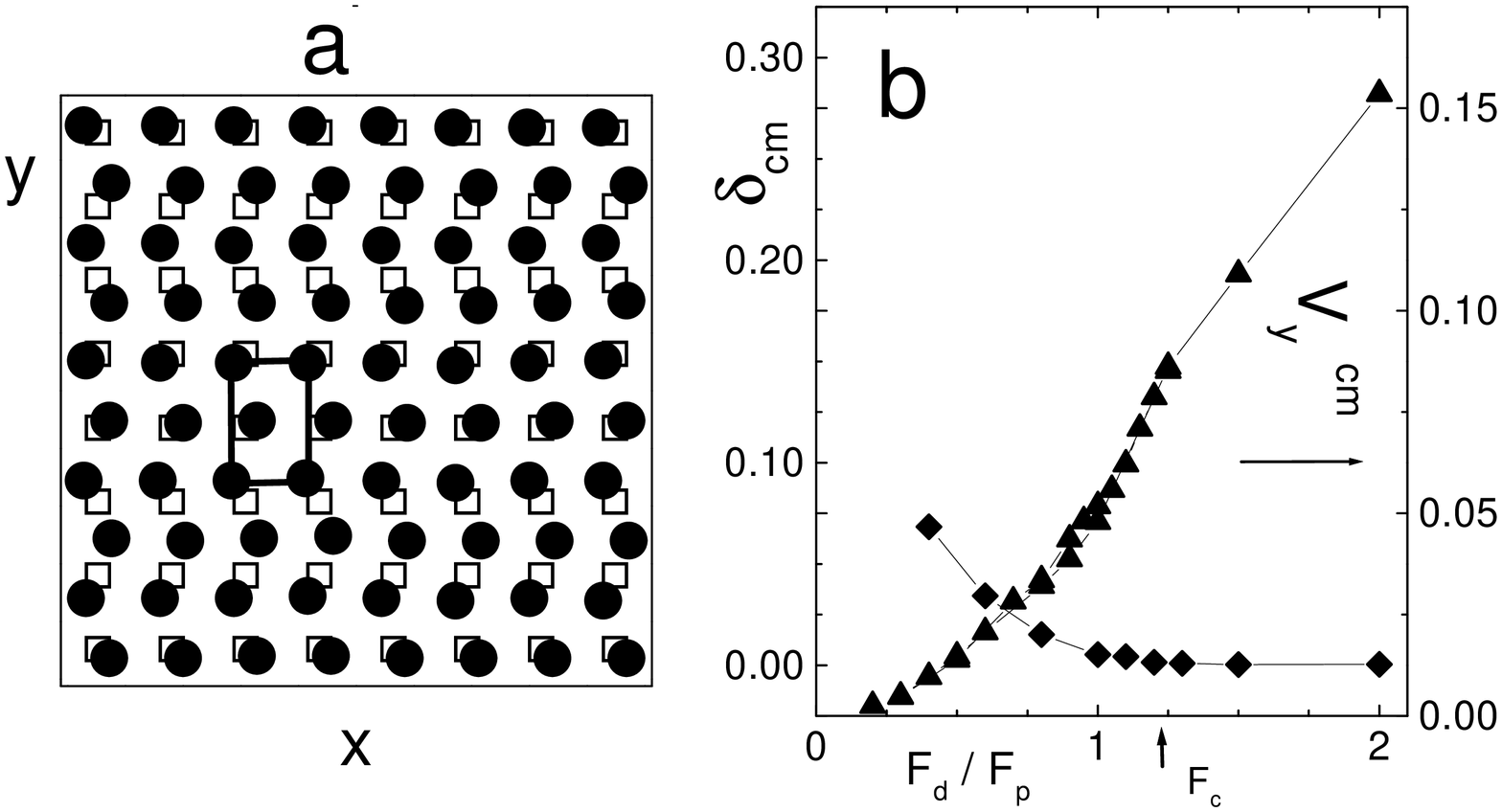,height=4.5cm,width=7.2cm,clip=}
\caption { $B=1.25B_{\phi}$: a) Full circles: vortex
positions for infinite-drive state. Full line: VL primitive unit cell.
Open squares: CDL. b) CM velocity $V^{cm}_y$ (units
$d/\tau$) and $\delta_{cm}=\Delta V^{cm}_y/V^{cm}_y$. } 
\label{fig:f2n}
\end{figure}

The possible infinite-drive states and the corresponding $T_m$
are determined by equilibrium Monte  
Carlo simulations of the model described above, with  the vortex-CDL 
interaction potential replaced by its average in the $y$-direction. This
is a periodic washboard potential, consisting of grooves  of width
$2R_{cd}$ running along the  $y$-direction and separated by 
$a_{cd}$ along the $x$-direction. Inside the groove the vortex is
attracted to its center by a force of constant modulus $F<F_p$
in the $x$-direction. For $B>B_{\phi}$ these states have only two
possible spatial orders:  incommensurate  (essentially triangular) at
$B$ sufficiently large, and commensurate with the $x$-direction
periodicity at lower $B$. The precise $B$ value at which the
ground-state changes from one type to the other depends on a
non-trivial way on the model parameters. It is found that, for the
above described model parameters, two typical
such infinite-drive states 
are the  vortex-lattices (VL) for $B=2B_{\phi}$, with the $P({\bf r})$
shown in Fig.\ \ref{fig:f1nn}.b.1, and for $B=1.25B_{\phi}$ with the vortex
arrangement shown in Fig.\ \ref{fig:f2n}.a. The latter can also be seen as
consisting of pairs of  vortex chains located within the grooves
and displaced relative to one another by half intra chain spacing. 
It is found that the VL for $B=2B_{\phi}$ melts to a vortex liquid at
$T_m=0.9J$, and that the VL for $B=1.25B_{\phi}$ first melts
to a smectic phase at $T/T_{sm}\sim 0.85J$ and then to a liquid at much
higher $T$. The results reported here for $B=1.25B_{\phi}$
 are restricted to $T/T_{sm}\sim 0.1$.
Monte Carlo simulations are also used to determine
the zero-drive states for these $B$ values. For
$B=2B_{\phi}$ a VL commensurate with the CDL and having one extra vortex at
the center of the CDL primitive unit cell is obtained, in agreement
with previous results \cite{nor,wmgc}. This state's $P({\bf r})$ is
shown in Fig.\ \ref{fig:f1nn}.b.4. For $B=1.25B_{\phi}$ a complex commensurate
VL with several vortices in the unit cell is found.  

\begin{figure}
\epsfig{file=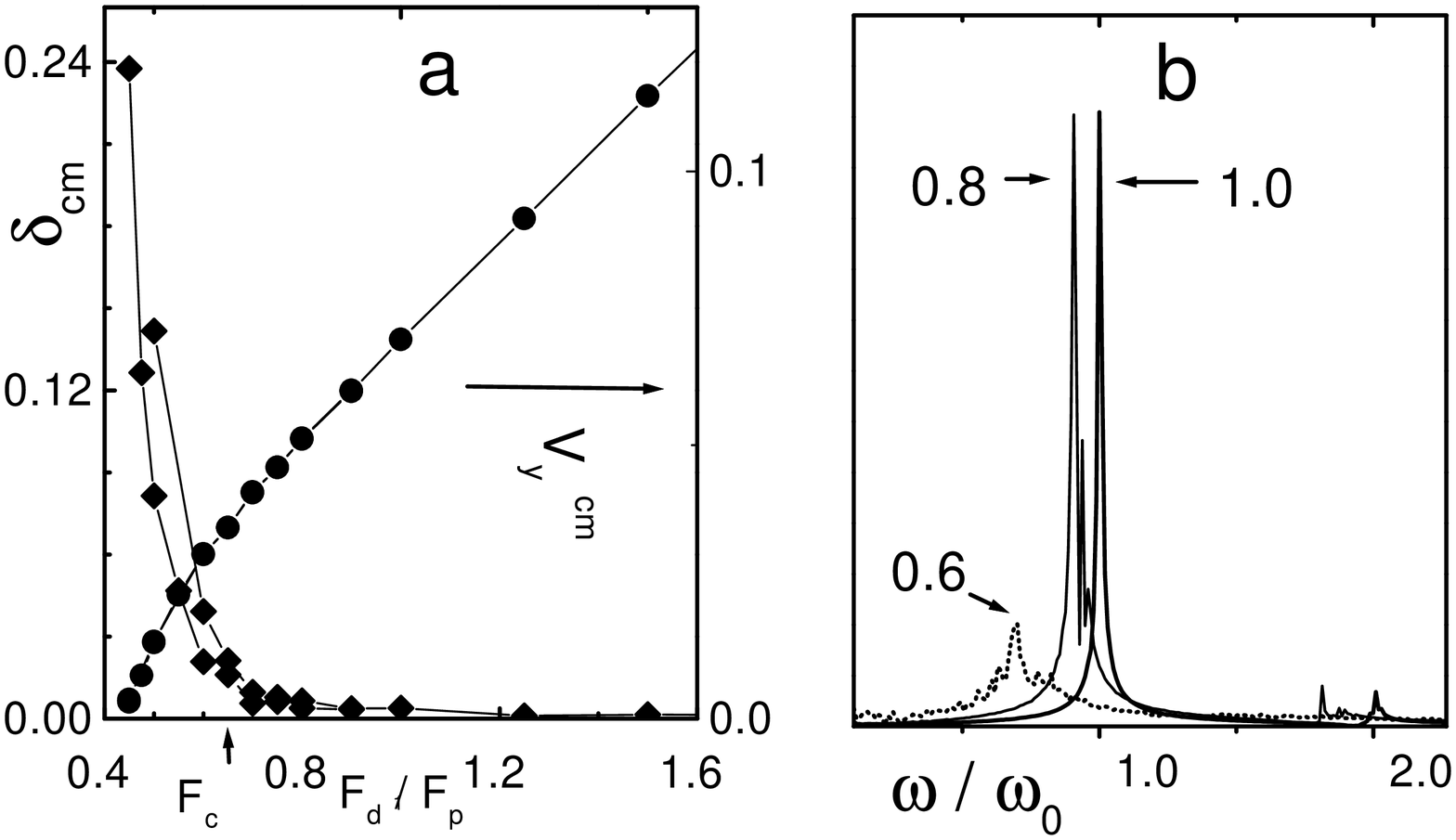,height=5cm,width=6.6cm,clip=}\caption{ $B=2B_{\phi}$: a) CM velocity $V^{cm}_y$(units  $d/\tau$) and
$\delta_{cm}=\Delta V^{cm}_y/V^{cm}_y$. b)
Correlation function  
$C^{sf}_y(\omega)$ at $F_d/F_p$ values indicated in
the figure. The ordinate axis (arbitrary units) in each curve is scaled
to show all curves in the same graph. $\omega_0$ denotes $\omega_{cm}$ at
$F_d/F_p=1.0$. Peaks are centered at $\omega_{cm}$ in all curves.} 
\label{fig:f3n}
\end{figure}

\begin{figure}
\epsfig{file=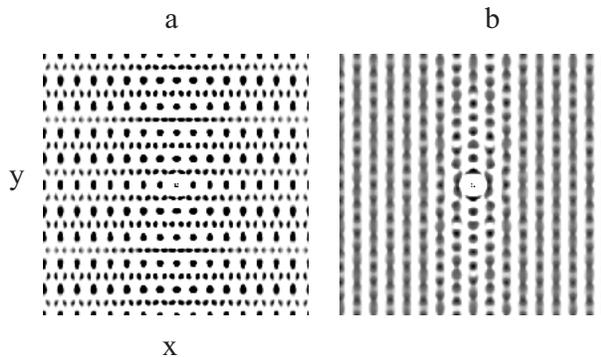,height=6.5cm,width=8.6cm,clip=}
\caption{ $P({\bf r})$ for $B=1.25B_{\phi}$ in $512\times 512$ space
lattice: a) elastic flow phase at $F_d=1.3F_p$, b) plastic flow phase at
$F_d=1.2F_p$.}    
\label{fig:f4n}
\end{figure}

The results of the dynamical simulations carried out here, to be
discussed in detail next, identify two  dynamical phases,  
elastic flow  and plastic flow, and a  transition between them
around  $F_d=F_c(T)$. It is found that for  $B/B_{\phi}=1.25$ and
$T/T_m\sim 0.2$,  $F_c/F_p \sim 1.25$ and that   
for  $B/B_{\phi}= 2$ the $F_c(T)$ curve is that shown in Fig.\
\ref{fig:f1nn}.a.  In both flow regimes the 
time-averaged velocity of all vortices is in the direction of drive.
The  $V^{cm}_y( F_d) $  curves ( V-I
curve ) at low $T$ are  depicted in Figs.\ \ref{fig:f2n}.b, \ref{fig:f3n}.a. 
The detailed properties of these phases are as follows.

i){\it Elastic flow: $F_d > F_c$, low $T$,  both $B$}. The moving
vortices remain  localized with respect to each other around the
infinite-drive state  
relative equilibrium positions. This is suggested by  $P({\bf r})$
shown in Figs.\ \ref{fig:f1nn}.b.1, \ref{fig:f1nn}.b.2 and \ref{fig:f4n}.a.
These consist of isolated spots with the  
symmetry of this state. As $F_d \rightarrow F_c$ these spots
become larger, indicating increase in relative vortex motion in the CM
frame. The   
time-averaged   velocity of all vortices is, within the simulation  errors,
$V^{cm}_y$, since $\Delta V^{cm}_{\alpha}/V^{cm}_y$ (${\alpha}=x,y$) is
small  for $F_d > F_c$ ( Figs.\ \ref{fig:f2n}.b, and \ref{fig:f3n}.a). Similar
results are found for $\Delta V^{cm}_x/V^{cm}_y$. 
Vortex motion consists of a translation in the direction of drive with
$V^{cm}_y$ and of periodic oscillations with frequency $\omega_{cm}=2\pi
V^{cm}_y/a_{cd}$ and higher
harmonics around the infinite-drive state equilibrium positions.
These oscillations are suggested by $C^{cm}_{\alpha}(\omega)$ and
$C^{sf}_{\alpha}(\omega)$  which are found to have sharp
peaks at $\omega_{cm}$ and peaks with smaller amplitudes at some harmonics 
as illustrated in Fig.\ \ref{fig:f3n}.b.  This frequency coincides with
the oscillation frequency of the vortex-CDL interaction in the frame
moving with  the CM. This suggests that the
periodic vortex motion is the (non-linear) elastic response of the VL
to this interaction. The frequency dependence of
$C^{cm}_{\alpha}(\omega)$ and $C^{sf}_{\alpha}(\omega)$ suggest 
long-range time order.  

ii){\it Transition region: $F_d \sim F_c$, low $T$, both $B$}. Vortex
spatial order changes in a small $F_d$ interval. This is seen  both in
$P({\bf r})$  and in  $S({\bf k})$, for ${\bf k}$ equal to high-drive
ground-state reciprocal lattice vectors, (Fig.\ \ref{fig:f1nn}). The
value of $F_c$ is estimated as the mid point of this interval, as
indicated in Fig.\ \ref{fig:f1nn}. This rapid change in the vortex spatial
order suggests a discontinuous jump in $S({\bf k})$. However, the
present data  cannot rule out a sharp crossover.
In the same $F_d$ interval, the $V^{cm}_y\times
F_d$ curves  changes slope, as seen in  (Figs.\ \ref{fig:f2n}.b,
\ref{fig:f3n}.a) and the sharp peaks in $C^{cm}_{\alpha}(\omega)$ and
$C^{sf}_{\alpha}(\omega)$ are found to disappear.   

iii){\it Plastic flow: $F_d < F_c$, low $T$, $B=2B_{\phi}$}.
Plastic flow sets in for $F_d$ just below $F_c$ and relative motion
increases as $F_d$ decreases, as evidenced by the sharp change in slope
and continuous growth in $\Delta V^{cm}_y/V^{cm}_y$ shown in Fig.\
\ref{fig:f3n}.a. The vortices become pinned in the zero-drive state for
$F_d<0.45F_p$. It is found that all vortices become pinned
simultaneously at  $F_d= 0.45F_p$. The vortices are disordered for
$F_d$ just below  $F_c$ (Fig.\ \ref{fig:f1nn}.b.3) and,    
as $F_d$ further decreases, the vortices show increasing order in the 
zero-drive state symmetry,    as suggested by the $S({\bf
k})$ vs. $F_d$ curves for ${\bf k}$ equal to zero-drive state
reciprocal lattice vectors (Fig.\ \ref{fig:f1nn}.a). This increase in
order arise because as $F_d$ deacreses the vortices spend more of
their travel time near the zero-drive state equilibrium  positions.
There is no long range time order as shown by the absence of sharp peaks
in the frequency dependence of $C^{cm}_{\alpha}(\omega)$ and
$C^{sf}_{\alpha}(\omega)$ ((Fig.\ \ref{fig:f3n}.b)

iv){\it Plastic flow: low $T$, $F_d < F_c$, $B=1.25B_{\phi}$}.
For $F_d$ just below $F_c$ some  spatial order
remains. For  $x=0$ and $x=a_{cd}$, 
$P({\bf r})$  (Fig.\ \ref{fig:f4n}.b) consists of isolated spots,
indicating that  the double  vortex chain-structure of the infinite-drive state
transforms to a single chain one with, roughly, nearest neighbor chains 
displaced relative to each other by half intra chain  spacing, 
and that  the vortices oscillate around the
corresponding equilibrium positions. For larger $x$ the spots become
interconnected by  stripes in the direction of motion, indicating that
inter chain relative motion at these separations, and thus plastic
flow, is taking place.  
The behavior of $\Delta  V^{cm}_y/V^{cm}_y$ across the transition
(Fig.\ \ref{fig:f2n}.b)  shows little change, suggesting that plastic flow
is weak.  The frequency dependence of $C^{cm}_{\alpha}(\omega)$ and
$C^{sf}_{\alpha}(\omega)$ is found to remain sharply peaked. These results  
are interpreted as indicating that vortices move together with the
chain, oscillating around their equilibrium position
within it, but relative motion between chains is taking
place. As $F_d$ decreases further plastic flow continues with
increasing $\Delta V^{cm}_y/V^{cm}_y$, but the vortices 
remain ordered in single chains and are unpinned for $F_d>0$ (Fig.\
\ref{fig:f2n}.b). It is expected that vortex reordering and pinning
would take place as $F_d$ decreases. The failure to see this in the
present simulations  is interpreted as evidence that trapping in a
metastable state is taking place.

v){\it T-dependence}. As $T$ increases $F_c(T)$ also increases,
because thermal fluctuations make the  infinite-drive state 
VL softer.  Thermal fluctuations also lead to smaller values of 
$S({\bf k})$. This makes the dynamical transitions described above more
difficult to observe, particularly for  $B=1.25B_{\phi}$. For 
$B=2B_{\phi}$ the $F_c(T)$ curve is shown in Fig.\ \ref{fig:f1nn}.a.

For other values of $B>B_{\phi}$, and drive in the (0,1)or (1,0)-directions,
the possible infinite-drive states are either a nearly triangular VL, or
a VL with $n$ vortex-chains trapped in each  groove, so that a similar
dynamical phase diagram is anticipated.   

The experimental verification of the new dynamical phases and  of the
transition  predicted in this paper may be possible in clean  superconducting
films with sufficient thermal fluctuations to allow relaxation to the
infinite-drive state to take place at large drives. Signatures of these
phases  are present in the V-I curve and in the noise power spectrum.
Visualization of the spatial order may also be possible using
decoration or fast STM techniques.




\end{multicols}

\end{document}